**Title:** Switching Propulsion Mechanisms of Tubular Catalytic Micromotors

*Paul Wrede, Mariana Medina-Sánchez\*, Vladimir M. Fomin\*, Oliver G. Schmidt*

Dr. Mariana Medina-Sánchez, Paul Wrede, Prof. Vladimir M. Fomin, Prof. Oliver G. Schmidt
Institute for Integrative Nanosciences, Leibniz IFW Dresden
Helmholtzstr. 20, 01069 Dresden, Germany

Paul Wrede
Max-Planck-Institute for Intelligent Systems
Heisenbergstraße 3, 70569 Stuttgart, Germany

Prof. Vladimir M. Fomin
National Research Nuclear University MEPhI (Moscow Engineering Physics Institute),
Kashirskoe shosse 31, 115409 Moscow, Russia

Prof. Oliver G. Schmidt
Center for Materials, Architectures and Integration of Nanomembranes (MAIN),
TU Chemnitz, Reichenhainer Strasse 10, 09107 Chemnitz, Germany
E-mail: m.medina.sanchez@ifw-dresden.de, v.fomin@ifw-dresden.de

Different propulsion mechanisms have been suggested for describing the motion of a variety of chemical micromotors, including the bubble-recoil mechanism, which has attracted great attention in the last decades due to its high efficiency and thrust force, enabling several applications in the fields of environmental remediation and biomedicine. Bubble-induced motion has been modeled including three different phenomena: capillarity, bubble growth, and bubble expulsion. However, most of those models have been suggested independently based on a single influencing factor (i.e. viscosity), limiting the understanding of the overall micromotor performance. In this work, we study the combined influence of medium viscosity, surface tension and fuel concentration on the switching behavior between different propulsion mechanisms in the same micromotor. Furthermore, we propose a holistic theoretical model that explains the three propulsion mechanisms, obtaining good agreement with the recorded experimental data.

# 1. Introduction

Over more than a decade, a lot of efforts have been made to develop synthetic micro-devices or micromotors that address, for instance, targeted medical treatments[1–4] and environmental remediation.[5–7] Moreover, micromotors are interesting model systems for studying motion mechanisms as well as other physical phenomena at low Reynolds numbers.[8,9] In particular, bubble-propelled catalytic micromotors have been of broad interest, as they show excellent motion performance in terms of speed and thrust force with simple designs.[10–14] These micromotors are often asymmetrical tubes, which are fabricated by strain-induced self-rolling of nanomembranes,[10,11,15] template-based electrodeposition,[12] or two-photon laser lithography.[16] They are provided with an inner surface of a catalytic material (e.g. Pt, Ag, enzyme) and an outside functional layer (e.g. Au, $SiO_2$, $TiO_2$) for their guidance, in situ reactions or for biofunctionalization purposes.[17–22] A chemical reaction between the fuel liquid and a catalyst coated on the inside of the micromotor converts chemical energy into mechanical motion. The most commonly used reaction is the decomposition of $H_2O_2$ into $O_2$ and $H_2O$ using Pt as catalyst.[11,12,23] Bubbles are formed by oxygen produced inside the tube, leading to a uni- or bidirectional movement, depending on the micromotor geometry.[24]

The physics behind the bubble propulsion mechanism was identified first by Mei et al.,[10] and has started to be explored in more depth in subsequent years.[11,25–27] The bubble propulsion is a result of three phenomena: i) capillarity,[28] ii) bubble growth,[27] and iii) bubble expulsion[29], depicted in **Figure 1a**. The capillarity introduces a capillary force when the oxygen bubble touches the inner wall of the tube, which leads to a movement of the bubble through its lumen. While the bubble continues growing by collecting oxygen, a fuel flow is induced, propelling the micromotor against the direction of the bubble release. Nucleation of several bubbles inside the tube has also been observed during this process.[28] During bubble generation, the bubble grows due to the continuous production of oxygen at the opening of the tube, displacing both the surrounding fluid and the micromotor.[27] As soon as the bubble detaches from the opening of the tube, the jet-like propulsion mechanism emerges. The bubble expulsion results in a frequent release of bubbles from the tube giving rise to a fluid flow opposite to the direction of the bubble motion. The propulsion of the micromotor is then analyzed by conservation of momentum in the system engine/fluid/bubble.[29] Based on our observations, the same micromotor can experience different propulsion modes when varying the experimental conditions (e.g. surface tension, fuel concentration and viscosity).

We focus on the three aforementioned phenomena and show that different propulsion mechanisms can be observed individually or simultaneously in the same experiment **(Figure 1b)**. We fabricate 50 µm long hollow conical micromotors with a 10 µm radius of the larger opening and a semi-cone angle of 5° using two-photon lithography **(Figure 1c)**, which allows for the control of the geometrical parameters with high yield (95 %), reducing the variability of experimental data (maximum standard deviation of the resulting velocities for ca. 20 micromotors is 8%). Based on these structures, the influence of different concentrations of fuel and surfactant as well as of the viscosity of the environment on the propulsion mode is analyzed. This provides new insights into the propulsion mechanisms and the possibility to further optimize and simulate the behavior of bubble propelled micromotors for operating in more complex media. Although some of those parameters have been taken into account in previous works, [26,34–37] there are no universal geometric parameters of the swimmers studied in the literature, resulting in different values of the speed among the produced micromotors, complicating the realization of a universal model that describes more accurately the micromotor motion.

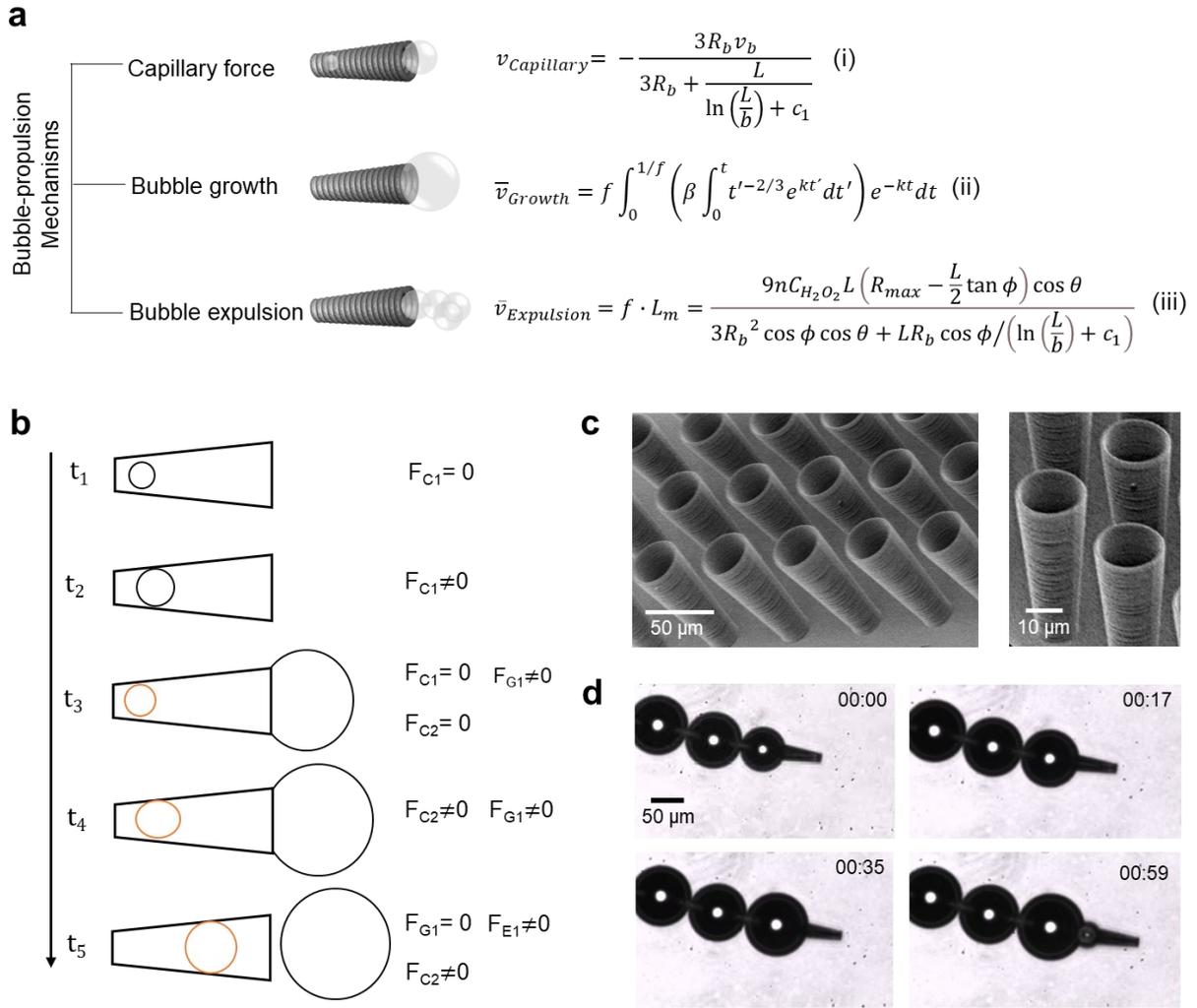

**Figure 1**. a) Schematic of the propulsion phenomena: i) capillarity [28], ii) bubble growth [27] and iii) bubble expulsion,[29] which can occur during the motion of a catalytic micromotor, and the corresponding micromotor average speed formulas. b) Illustration of the forces acting at different time intervals during the motion of a catalytic micromotor. c) SEM images of 50 μm long catalytic micromotors with an upper radius of 10 μm and a semi-cone angle of 5°. d) Snapshots of the micromotor motion in a solution containing 2.5% $H_2O_2$ and 5% SDS. The bubble growth and expulsion over time (in sec) are clearly observed. These photographs illustrate that during the motion of the micromotor, various propulsion mechanisms are dominant at different times.

## 2. Results

By changing the composition of the fuel, we alter the hydrodynamics of the micromotor, which directly influences the speed of the micromotor and switches between different propulsion modes. For instance, increasing fuel concentration (i.e. hydrogen peroxide, $H_2O_2$) leads to a higher production rate of oxygen, thus causing bubbles to grow faster and leave the tube more frequently, hence propelling the micromotor at higher speed. Alternatively, the use of surfactants (i.e. Sodium dodecyl sulfate, SDS) reduces the fluid surface tension, allowing the bubbles to move faster through the micromotor and detach easier from the larger tube opening. Another way to influence the motion of the micromotor is by changing the viscosity of the fuel by adding for example methylcellulose (MC) at different concentrations[38] as it influences the drag force acting on the micromotor. Thus, in this work, a switching between propulsion mechanisms depending on the concentrations of SDS, $H_2O_2$, and MC is thoughtfully analyzed. The impact of each parameter was quantified by measuring the bubble diameter, the bubble release frequency and the speed of the micromotor, and compared to the theoretical data obtained by applying each of the propulsion mechanisms under consideration.

### 2.1. Influence of the Surfactant Concentration

To investigate the influence of the surfactant concentration on the performance of the micromotors, SDS was added to the fuel liquid in different concentrations (1.25%, 2.5%, 5%, 7.5%, 10%, 20% and 30%), while maintaining a constant concentration of $H_2O_2$ (2.5%). According to Wang et al.,[35] anionic surfactants such as SDS have a stronger speed increasing effect than non-ionic or cationic ones. Consequently, SDS was used in these experiments to reduce the surface tension for reaching high speeds. Looking at the diameter of the bubbles in the individual micromotors in **Figure 2a and b**, the bubbles decreased in diameter with increasing concentration of the surfactant in the interval from 1.25% to 10% SDS, while bubbles grew much larger when the SDS concentration was increased to 20% and 30%. An opposite behaviour was found for the average bubble release frequency. From 1.25% to 10% SDS the average bubble release frequency increased, but it dramatically decreased for concentrations of 20% and 30% SDS. The micromotors stopped completely at an SDS concentration of 30% by forming only one big bubble. This behaviour was confirmed by the obtained average speed and trajectories of all observed micromotors (**Figure a, c**). The reduced speeds were caused by large bubble diameters, while higher speeds were linked to higher bubble release frequencies (**Figure 2b**). The speed of the micromotors increased from about 160 µm/s at 1.25% SDS almost linearly to about 350 µm/s at 10% SDS, while at 20% and 30% SDS, a rapid decrease in the speed of the micromotors was observed. An anionic surfactant, such as SDS, reduces the surface tension in the medium. This facilitates the formation and escape of oxygen bubbles from the tube, and the bubble diameter decreases as the bubbles are fed with less oxygen over a shorter time period. At 30% SDS, micromotor motion ceased as only one bubble was continuously growing at the larger opening of the micromotor. **Figure 2c** shows that none of the existing reported models predicts the micromotor speed accurately. The bubble growth mechanism is the one, which approaches the experimental data most closely, having a difference in the threshold micromotor speed of ca. 152 µm/s.

The effect of surfactant in the fuel solution was previously investigated by Wang et al.,[35] who reported that the speed of their catalytic micromotors no longer increased from the SDS concentration greater than 0.3%, differing from the threshold of 10% SDS observed in our work. These results were explained by the adsorption of the surfactant at the interfaces,[39] explaining the decrease in micromotor speed for SDS concentrations over the threshold value. This theory states that the absorption of the surfactant takes place in two stages. [40] The first stage is characterized by the chemical interaction of the surfactant molecules and the solid surface, which causes the surfactant molecules to be transferred onto the solid surface. In the second stage, the adsorbed molecules form a hemimicelle through the interaction of the molecules with each other. These two steps can be summarized with the two following equilibrium (**equations [1] and [2]**):[39]

$$\text{Site} + \text{Surfactant} \leftrightarrow \text{Adsorbed Surfactant} \qquad [1]^{[39]}$$

$$(n-1)\text{Surfactants} + \text{Adsorbed Surfactant} \leftrightarrow \text{Hemimicelle} \qquad [2]^{[39]}$$

The amount of surfactant adsorbed ($\Gamma$) at a concentration $C$ is described by the following isotherm equation:

$$\Gamma = \frac{\Gamma_\infty k_1 C(\frac{1}{n} + k_2 C^{n-1})}{1 + k_1 C(1 + +k_2 C^{n-1})} \qquad [3]^{[39]}$$

where $k_1$ and $k_2$ are the equilibrium constants for the chemical reactions, $\Gamma_\infty$ is the maximum amount of surfactant, which can be absorbed at high concentrations, and $n$ is the hemimicelle aggregation number. $\Gamma_\infty$ and $\Gamma$ can be measured experimentally.[39] There are certain limiting cases that lead to the Langmuir equation. If $k_2 \to 0$ and $n \to 1$, **equation [3]** changes to

$$\Gamma = \frac{\Gamma_\infty k_1 C}{1 + k_1 C} \qquad [4]$$

If $n > 1$ and $k_2 C^{n-1} \ll 1/n$, **equation [3]** acquires the form

$$\Gamma = \frac{(\Gamma_\infty/n)k_1 C}{1 + k_1 C} \qquad [5]$$

According to **equations [4] and [5]**, the amount of SDS onto the micromotor surface is a function of the SDS concentration, the aggregation number $n$ and the reaction constant $k_1$. The latter, in turn, depends on the amount of the material available for the reaction and thus on the geometry of the micromotor. Accordingly, there are different limits for the adsorption of surfactant depending on geometry and material of the micromotor leading to changes in the micromotor motion as the surface tension lowering effect occurs only when SDS is adsorbed. This could explain the different results of Wang et al. [35] and the present work by the fact that smaller micromotors had less material, which could adsorb the SDS. Consequently, the limit of adsorption is much less than for the micromotors used in the current work, reducing their speed earlier, at 0.3% SDS. This limit of adsorption is also the reason for the decrease of speed for surfactant concentrations over 10%, as it makes the fuel liquid more viscous since the material has already reached the adsorption limit. When reaching it, the increase in viscosity is dominant over the surface tension, reducing the micromotor speed.

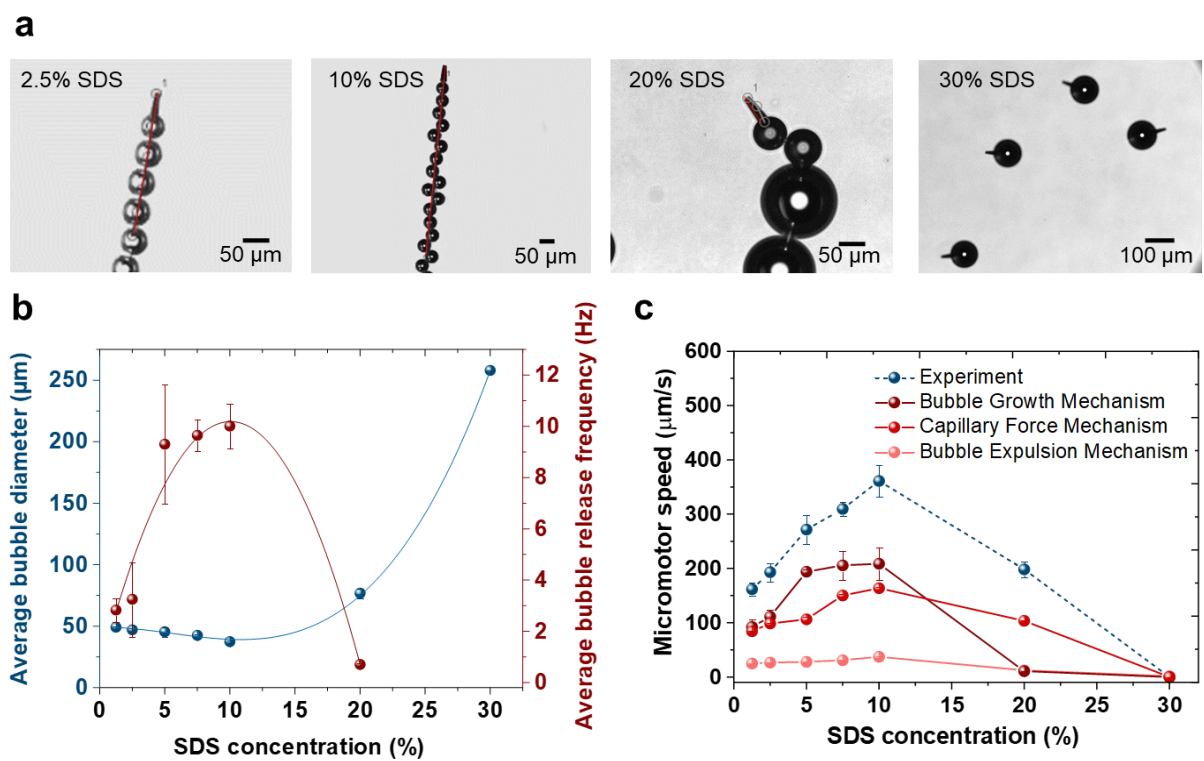

**Figure 2.** a) Micromotors trajectory and expelled bubble size under different SDS concentrations. Time for all videos was set to 1.15 s (**Video S1**). From 2.5% SDS to 10% SDS, the bubbles become smaller and the speed of the micromotor increases (as seen through a longer trajectory for 10 % SDS). In contrast, at the concentration of 20% SDS, the bubble is much larger and the micromotor is much slower compared to the image at 10% SDS. At 30% SDS, the surface tension is so high that the bubble just continues to grow and it is not able to detach from the micromotor surface. Consequently, the micromotor stopped This implies that the growth mechanism is dominant in the propulsion. b) Average bubble diameter and bubble release frequency as a function of SDS concentration (error bars correspond to 15-20 observed

micromotors). c) Comparison between the micromotor speed calculated for different propulsion mechanisms with the obtained experimental data.

## 2.2. Influence of the Hydrogen Peroxide Concentration

The energy required to propel catalytic micromotors is provided by the decomposition of the surrounding $H_2O_2$ fuel into oxygen gas, which is catalysed by the tube inner Pt layer. To investigate the influence of $H_2O_2$ concentration, different fuel liquids with $H_2O_2$ concentrations (2.5%, 3.75%, 5%, 7.5%, 10%, 20% and 30%) and a constant SDS concentration of 5% were used. In **Figure 3a and b**, the bubbles of individual micromotors decrease in diameter with increasing $H_2O_2$ concentration within the whole range under consideration. This is different from the behaviour of the bubble release frequency, which increases with increasing $H_2O_2$ concentration (**Figure 3b**). We confirmed this effect by measuring a statistically relevant number of around 15-20 catalytic micromotors. The average speed increased linearly from about 250 μm/s at 2.5% $H_2O_2$ to about 700 μm/s at 10% $H_2O_2$. At higher $H_2O_2$ concentrations, more $O_2$ can be produced by chemical decomposition, underpinning these results (**Figure 3c**). Accordingly, the bubble release frequency increases, and so does the speed of micromotors. Due to this increased frequency and the associated shorter time, during which the bubbles remain in the tube, they can also collect less oxygen leading to a smaller increase in speed for $H_2O_2$ concentrations over 10%. Since the amount of oxygen inside the bubble and the diameter of the bubble are directly interrelated, the bubble becomes smaller with increasing concentration of $H_2O_2$. **Figure 3c** shows that none of the existing mechanisms predicts the micromotor speed precisely enough as when varying SDS concentration. The only exception is the expulsion propulsion model for the speed of the micromotors at 30% $H_2O_2$ (details explained in sections 2.4 and 2.5). The catalytic micromotors of Solovev, et al. [41] showed an increase of their speed with increasing concentration of $H_2O_2$ in the range of 2.5% to 10%. However, the conical catalytic micromotors used here are four to five times faster than those of Solovev et al., which might be caused by the different fabrication methods[41]. It could be confirmed that the increase of speed by increasing the $H_2O_2$ concentration over 20% became much less than for concentrations below this threshold. The proposed decrease of the speed at very high concentrations of $H_2O_2$ could not be observed during the experiments performed in the present approach. Anyway, this behaviour could occur at still higher concentrations. The reason for this phenomenon is the switching between propulsion mechanisms, which will be discussed in sections 2.4 and 2.5 below. Especially at concentrations of 30% $H_2O_2$, the micromotors tend to stop soon after the injection of the fuel solution, probably caused by the bubble blockage (where more bubbles are formed inside the tube than can be released, which could cause a bubble to be pushed outside the smaller opening of the tube and to block it) proposed by Klingner et al.[28]

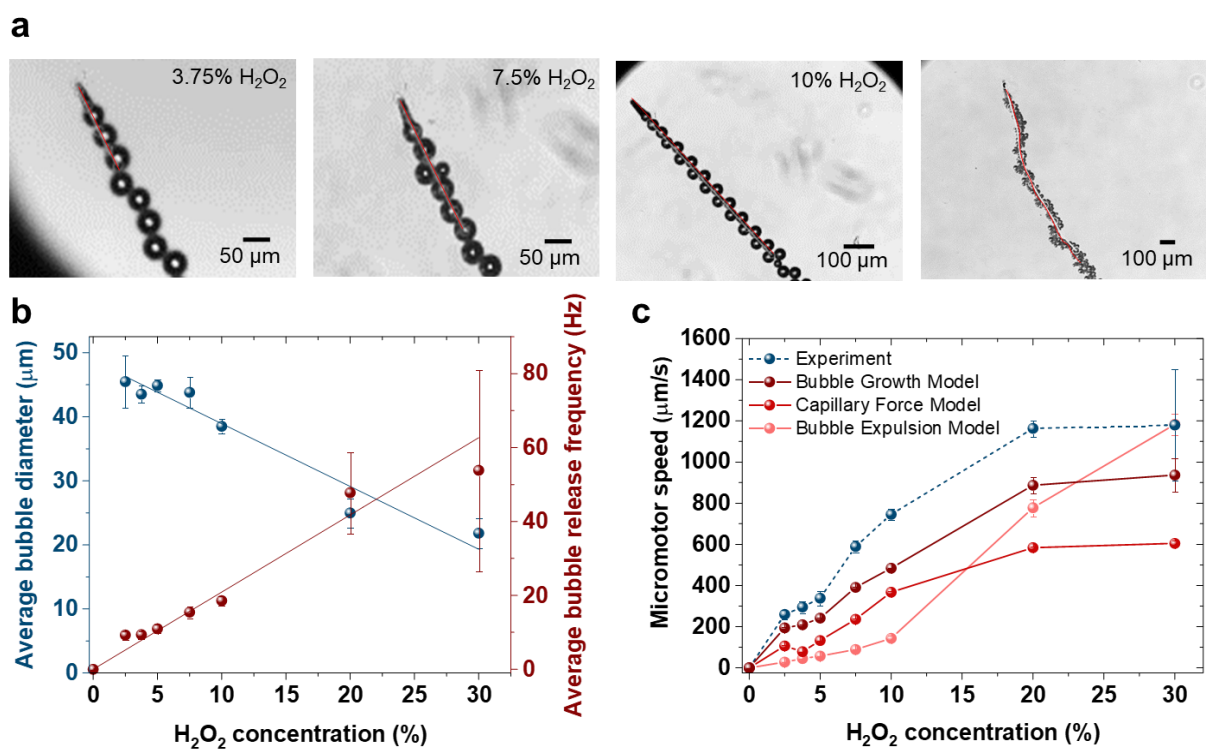

**Figure 3.** a) Trajectories of micromotors captured from the experiments with different $H_2O_2$ concentrations. The speed is increasing with higher concentrations, which is shown by longer trajectories for the same time interval of 0.29 s (**Video S2**). b) Bubble diameter and bubble release frequency as a function of the $H_2O_2$ concentration. The bubble diameter decreases

dramatically with increasing $H_2O_2$ concentration from 10% to 30% leading to switching of propulsion mechanisms from a combined one to the expulsion mechanism. The switching occurs as long as the bubble diameter is nearly as large as the diameter of the micromotor larger opening. c) Speed of micromotors as a function of the $H_2O_2$ concentration and comparison with the speed calculated for the three propulsion mechanisms.

### 2.3. Influence of the Fluid Viscosity

The medium surrounding the catalytic micromotor also has a major influence on its performance. According to Newton's third law of motion, the driving force $\boldsymbol{F_j}$ that moves the micromotor forward is counteracted by a force acting against the direction of motion. The latter force that slows down the micromotor is the drag force $\boldsymbol{F_v}$ of the medium, which is directly proportional to its viscosity. This influence is investigated within the present study by adding MC to the fuel solution. MC increases the viscoelastic properties of the medium. A number of values of MC concentration (0%, 0.05%, 0.1%, 0.15%, 0.2%, 0.25%, 0.4% and 0.6%), corresponding to fluid viscosities of 0.0013, 0.0018, 0.002, 0.0022, 0.0025, 0.0028, 0.004 and 0.006 Pa·s (measurements done with a conventional rheometer), respectively, were used in this experiment. The $H_2O_2$ concentration was kept constant at 2.5% and the SDS concentration was kept at 5%, offering the best motion stability.

As seen in **Figure 4a and b**, the bubbles became larger with increasing MC concentration, while the speed and the bubble release frequency decreased. The decrease of the bubble release frequency slowed down the micromotors. In **Figure 4c** the trend of a lower speed due to a higher viscosity was confirmed for all observed micromotors (15-20 for each data point). The graphs for the simulated speed of the micromotors for the three considered propulsion mechanisms confirm once again that none mechanism alone is sufficient to interpret the micromotor speed, especially at lower viscosities. The reason for the observations is the passivating effect of MC on platinum inside the tubes. This results in a reduction of the chemical activity of platinum, which, in turn, leads to a reduced decomposition of $H_2O_2$ and thus to a weaker $O_2$ production. As a consequence, the bubble release frequency goes down, allowing the bubbles to stay longer inside the tube and collect more $O_2$, resulting in an increased bubble diameter. Therefore, bubbles grow for longer time inside the micromotor, blocking it and hindering its propulsion. The influence of the viscosity of the medium on the performance of micromotors was already investigated by Wang et al.[36] Their results were based on mathematical modelling of the average micromotor speed.

Wang et al. [36] also concluded that the higher the viscosity of the medium, the lower the speed of micromotors. However, the simulated average speed for the micromotors decreased faster than it does for the micromotors investigated within the present approach. This fact could be caused by the different geometric parameters as mentioned at the beginning of the present section.

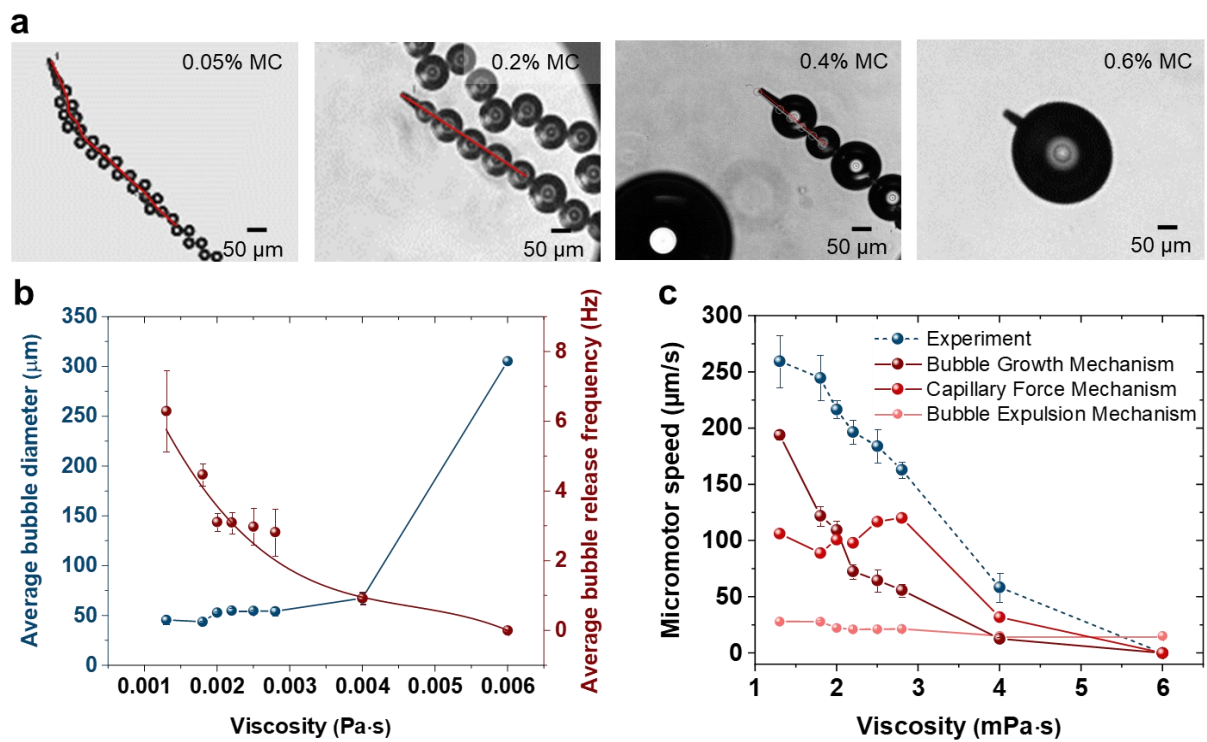

**Figure 4.** a) Trajectories taken from the experiments with varying MC concentration. The speed of the micromotor decreases with rising concentration of MC, as the latter increases the viscosity of the medium. At the concentration of 0.6% MC, only a bubble growth is present

(Video S3). b) Average bubble diameter and expelling frequency as a function of the SDS concentration. c) Speed of micromotors as a function of viscosity of the fuel solution and comparison of our experimental data with the micromotor speed obtained using different propulsion mechanisms. Trajectory recorded during a time interval of 1.3 s.

### 2.4. Combining regimes of propulsion

The experiments show that for the used micromotors all three regimes of propulsion are influenced by viscosity, surface tension and fuel concentration, which vary the bubble diameter and the bubble release frequency, leading to a change in the micromotor speed and the dominant propulsion mode. As shown in the present section, an adequate theoretical description of the speed of the micromotors requires a combination of the three mechanisms of propulsion. According to Newton's third law of motion and considering a steady motion at very low Reynolds numbers, the drag force $\boldsymbol{F}_{Drag}$ at an instant $t$ compensates the driving force $\boldsymbol{F}_{Driving}$ acting on the micromotor:

$$\boldsymbol{F}_{Drag}(t) = \boldsymbol{F}_{Driving}(t) \qquad [6]$$

The drag force acting on the micromotor depends on its instantaneous speed $v(t)$, the length $L$, the maximal radius $R_{max}$, the semi-cone angle $\phi$ and the fluid viscosity $\eta$:

$$\boldsymbol{F}_{Drag}(t) = \frac{2\pi\eta L v(t)}{\ln\left(\frac{L}{b}\right) + c_1} \qquad [7]$$

Here $c_1$ is defined by the geometry of the tube

$$c_1 = -\frac{1}{2} + \ln 2 - \frac{2-\xi\tan\phi}{2\xi\tan\phi}\left[\frac{2}{2-\xi\tan\phi}\ln\left(\frac{2}{2-\xi\tan\phi}\right) - \frac{2-2\xi\tan\phi}{2-\xi\tan\phi}\ln\left(\frac{2-2\xi\tan\phi}{2-\xi\tan\phi}\right)\right]$$

with $b = R_{max} - \frac{L}{2}\tan\phi$ and $\xi = \frac{L}{R_{max}}$ [51].

$F_{Driving}$ depends on the propulsion mechanism at a given instance. For the capillary force mechanism,[28] a driving force is a capillary force induced by the bubble motion through the tube. Accordingly, the driving force as a function of time $F_C(t)$ depends on the instantaneous bubble radius $R_b(t)$ and the absolute speed of the bubble with respect to the vessel, which is given by $v_b(t) + v(t)$, $v_b(t)$ being the instantaneous speed of the bubble relative to the tube:

$$F_C(t) = 6\pi\eta R_b(t)(v_b(t) + v(t)).  \quad [8]$$

Starting from $m_b(t) = G_{O_2} t = \frac{4}{3}\rho_{O_2}\pi R_b(t)^3$ we find the maximal bubble radius $R_b\left(\frac{1}{f}\right) = \sqrt[3]{\frac{3G_{O_2}}{4f\pi\rho_{O_2}}}$, where $f$ is the bubble release frequency, $\rho_{O_2}$ is the mass density of oxygen and $G_{O_2}$ is the rate of oxygen production.

The driving force within the bubble growth mechanism [27] $F_G(t)$ depends on the size of the bubble and its growth rate and therefore on the rate of oxygen production:

$$F_G(t) = \rho_w \pi R_b(t)^2 \left(\frac{3}{2}\right) C_s \dot{R}_b(t)^2 + R_b(t)\ddot{R}_b(t) \quad [9]$$

Here $C_s$ is a fitting constant taken from experimental data.

The driving force for the bubble expulsion mechanism[29] originates due to bubbles leaving the micromotor frequently, giving rise to a flow of the liquid against the directed motion of the bubbles. It is expressed by the following equation:

$$F_E(t) = 6\pi\eta R_b(t) v_b(t) \cos\theta, \quad [10]$$

where $\theta$ is the inclination angle of the bubble speed with respect to the inner wall of the tube. During the experiments, two bubbles inside the tube could be observed in accordance with the observations by Klingner et al.[28] Therefore, we considered a system consisting of the engine+fluid+two bubbles (**Figure 1b**). At the first stage, a bubble started to grow inside the tube. By collecting more oxygen, it grew larger until it touched the inner wall of the micromotor. At this instant, $F_C$ began to move the micromotor against the direction of the bubble motion.

When reaching the larger opening of the tube, the bubble grew further, being adhered to the opening. Meanwhile, a second bubble started to grow inside the micromotor, not touching the inner wall yet. At this stage, $\boldsymbol{F}_{Driving}(t)$ was only given by $\boldsymbol{F}_G(t)$ of the first bubble. At the next stage, the second bubble touched the inner wall of the tube, which resulted in $\boldsymbol{F}_{Driving}(t)$ represented as a sum of $\boldsymbol{F}_G(t)$ of the first bubble and $\boldsymbol{F}_C(t)$ of the second bubble. When the first bubble detached from the tube opening, we got the driving force as a sum of $\boldsymbol{F}_C(t)$ from the second bubble and $\boldsymbol{F}_E(t)$ due to the first bubble.

In summary, the overall driving force is a sum of all driving forces corresponding to the three propulsion modes. On various time intervals, different propulsion mechanisms are dominant, while some of the mechanisms may be not in force.

The micromotorspeed for each mechanism is calculated according to **Equation [7]**:

$$\boldsymbol{v}_{Capillary} = -\frac{3R_b v_b}{3R_b + \frac{L}{\ln\left(\frac{L}{b}\right) + c_1}} \qquad [11]^{[28]}$$

$$\bar{v}_{Growth} = f \int_0^{1/f} (\beta \int_0^t t'^{-2/3} e^{kt'} dt') e^{-kt} dt \qquad [12]^{[27]}$$

with $\beta = \gamma^4 \frac{\rho_w \pi}{9m_j}\left(\frac{3}{2}C_s - 2\right), \beta = \gamma^4 \frac{\rho_w \pi}{9m_j}\left(\frac{3}{2}C_s - 2\right), k = \frac{2\pi\eta L}{m_j\left[\ln\left(\frac{L}{b}\right) + c_1\right]},$

$\gamma = \left[-\frac{3D_{H_2O_2} C_{H_2O_2}^\infty R^2 M_{O_2}}{8\rho_{O_2} L}\left(\frac{1}{\cosh(\beta L)}\right) - 1\right]^{1/3}$, where $m_j$ is the mass of the micromotor, $C_s$ is an empirical constant set to 6.67 [42], $C_{H_2O_2}^\infty$ is the concentration of $H_2O_2$ at the bubble-sealed opening of the microtube and $D_{H_2O_2} = 1.43 \times 10^{-9}$ m²/s according to Manjare et al. [27]

$$\bar{v}_{Expulsion} = f \cdot L_m = \frac{9nC_{H_2O_2}L\left(R_{max} - \frac{L}{2}\tan\phi\right)\cos\theta}{3R_b^2 \cos\phi \cos\theta + LR_b \cos\phi/\left(\ln\left(\frac{L}{b}\right) + c_1\right)} \qquad [13]^{[29]}$$

Here $C_{H_2O_2}$ and $n$ are the concentration of $H_2O_2$, and a parameter taken from the experiment, correspondingly. According to **Equations [12]-[14]**, the micromotor speed vanishes as soon as either $v_b$ or $f$ become 0. The time average micromotor speed in [13] and [14] is calculated by integration of the instantaneous speed over the period $\Delta t = \frac{1}{f}$ between two consecutive ejections divided by this period,

$$\overline{v} = f \int_0^{1/f} v(t)\, dt. \tag{14}$$

**Simulation and switching of propulsion mechanisms**

In the present section, we use the theoretical model in order to interpret the experimentally observed micromotor dynamics. For simulations based on **Equations [12]-[15]**, both parameters determined from our experiments and those taken from the literature are used (see **Table 1**).

**Table 1.** Geometric and materials parameters used for simulation of catalytic micromotors.

| Parameter | Value | Parameter | Value | Reference |
|---|---|---|---|---|
| $R_{max}$ [μm] | 10 | $\rho_{O_2}$ [kg·m$^{-3}$] | 1.429 | [1] |
| $R_{min}$ [μm] | 5.6 | $\rho_{Water}$ [kg·m$^{-3}$] | 998.2067 | [3] |
| $L$ [μm] | 50 | $\theta$ [°] | 0.001 | [29] |
| $\phi$ [°] | 5 | | | |

Our experiments imply that, depending on the composition of the fuel fluid, a switching of propulsion mechanisms occurs.

A dominant influence of the bubble growth mechanism could be observed for concentrations higher than 20% SDS (at 2.5% $H_2O_2$) as well as for concentrations of 0.4% MC and higher (with viscosity above 0.004 Pa·s) at 2.5% $H_2O_2$ and 5% SDS (**Figure 5a, b**). In those cases, not enough force is exerted on the bubble to detach it from the tube. This caused the bubble to keep on growing becoming much bigger than the larger opening of the micromotor (**Figures 2 and 4**). Due to the motion of the bubble in the micromotor, a relatively small contribution of the capillary force is also present, which can be neglected for a sufficiently long time because no second bubble formed inside the micromotor. An exclusive occurrence of the capillary force mechanism is not possible, because the bubble always leaves the interior of the micromotor within this combined regime. The dominance of the capillary force mechanism should be feasible if the bubble is kept inside the tube. Our experimental data suggest that starting from the concentration of about 20% $H_2O_2$ (the specific value depends on the system under study), the bubble expulsion mechanism dominates over the other two mechanisms. The reason for this is twofold. Firstly, the bubbles move through the tube so quickly, that their radii remain smaller than that of the tube (**Figure 3**). Accordingly, there is no contact between the bubble and the inner wall to induce a capillary force. Secondly, due to their small size, the bubbles leave the micromotor without adhering, so that no growth force emerges. This explains a flatter increase of the speed of the micromotors in the range of the $H_2O_2$ concentrations between 20% and 30% compared to the $H_2O_2$ concentration range between 2.5% and 10%, which is seen in **Figure 3**. When the concentration is as low as 10% $H_2O_2$ or lower, the influence of the other two propulsion mechanisms diminishes. As a result, this speed, as compared to the overall speed by the other two mechanisms, becomes progressively smaller. The higher speed induced by the bubble expulsion mechanism is not sufficient to completely compensate this trend leading to a lower overall increase of speed with increasing $H_2O_2$ concentration.

---

[1] *GESTIS chemicals database of the IFA, retrieved on 25.06.2019*

The instantaneous speed of the micromotors at 5% SDS, 20% $H_2O_2$, and 0.4% MC, respectively, show the switching of the propulsion mechanisms **(Figure 5b)**. For instance, for 5% SDS solution **(Figure 5b, i),** all mechanisms contribute to the micromotor motion: each quasi-oscillation represents one cycle of bubble dynamics that includes growth inside the tube, motion along it, and finally expulsion. At the same time, the micromotor speed increases to a peak value which is much higher than the average speed, caused by the bubble expulsion. Afterwards, the micromotor decelerates due to the drag force leading to a speed reduction compared to the average speed. As seen in **Figure 5b, i**, this deceleration becomes smaller at the point when a new bubble completes its motion along the tube and starts growing outside it. As this bubble has to move through the tube, also a capillary force emerges. Therefore, the stage between the peak value and the local minimum value of the micromotor speed, corresponds to the time interval in which the bubble growth and the capillary force mainly contribute to the micromotor speed. As a result, the average micromotor speed in the 5% SDS solution is achieved by a concerted action of the capillary force, bubble growth and expulsion. Starting from the 20% $H_2O_2$ concentration, the bubble expulsion is dominant. For 20% $H_2O_2$ solution (**Figure 5b, ii**), frequent small fluctuations around the average speed value of ~ 1200 µm/s are observed, caused by the high-frequency expulsion of bubbles. Therefore, the bubble expulsion mechanism plays a dominant role in this situation, when compared with the combination of the bubble growth and the capillary force mechanisms. In contrast, for 0.4% MC solution, which possesses a viscosity of ca. 4 mPa·s (**Figure 5b, iii**), a slow deceleration of the micromotor takes place over the whole time-interval of the measurement. The initial stage represents the micromotor speed after expulsion of a bubble. Afterwards a new bubble moves through the tube and starts to grow outside, though it is still attached, resulting in the same behavior as explained above for a micromotor moving in 5% SDS solution. Since the time of the bubble growth is much longer for a micromotor in a more viscous medium, the effect of the bubble expulsion on the micromotor speed could be neglected. In this case, the micromotor speed could be explained by a combination of the bubble growth and the capillary force mechanisms.

In summary, switching of the propulsion regime occurs by virtue of the following conditions:

$$F_{Driving} = \begin{cases} F_C, \text{if the bubble stays inside the tube;} \\ F_G \text{ for } R_b(t) > R_{max}, \text{when the bubble remains attached to the tube;} \\ F_E, \text{when the bubble has left the tube.} \end{cases} \quad [15]$$

For our simulations, all three mechanisms are considered, in particular for $H_2O_2$ concentrations lower than 20%, whereas for the $H_2O_2$ concentration above this value, only **Equation [13]** is used. Considering all mechanisms in our model, the simulated speed as a function of the $H_2O_2$ concentration (ranging from 2.5% to 30%) is in good agreement with the experimental data (**Figure 5a**). All three propulsion mechanisms are also considered for the whole range of MC concentrations (from 0.05 to 0.25%, corresponding to viscosities from 1.3 to 6 mPa·s, respectively), as well as for the whole range of SDS concentrations (from 1.25%, to 10%). As the growth mechanism is dominant for the MC concentrations of 0.4% and 0.6% as well as for the SDS concentrations of 20% and 30%, only **Equation [13]** is used. Moreover, it is worth mentioning, that the oxygen production rate $G_{O_2} = \frac{4}{3} f \rho_{O_2} \pi R_b^3 \left(\frac{1}{f}\right)$ is a cubic function of the maximal bubble radius. Since the bubble radius was not measured with high precision (in 2D by optical microscopy), deviations within the error bar strongly influence the calculated values in **Figures 2, 3, and 4**. This is the main reason of the variance between the theoretical and experimental data.

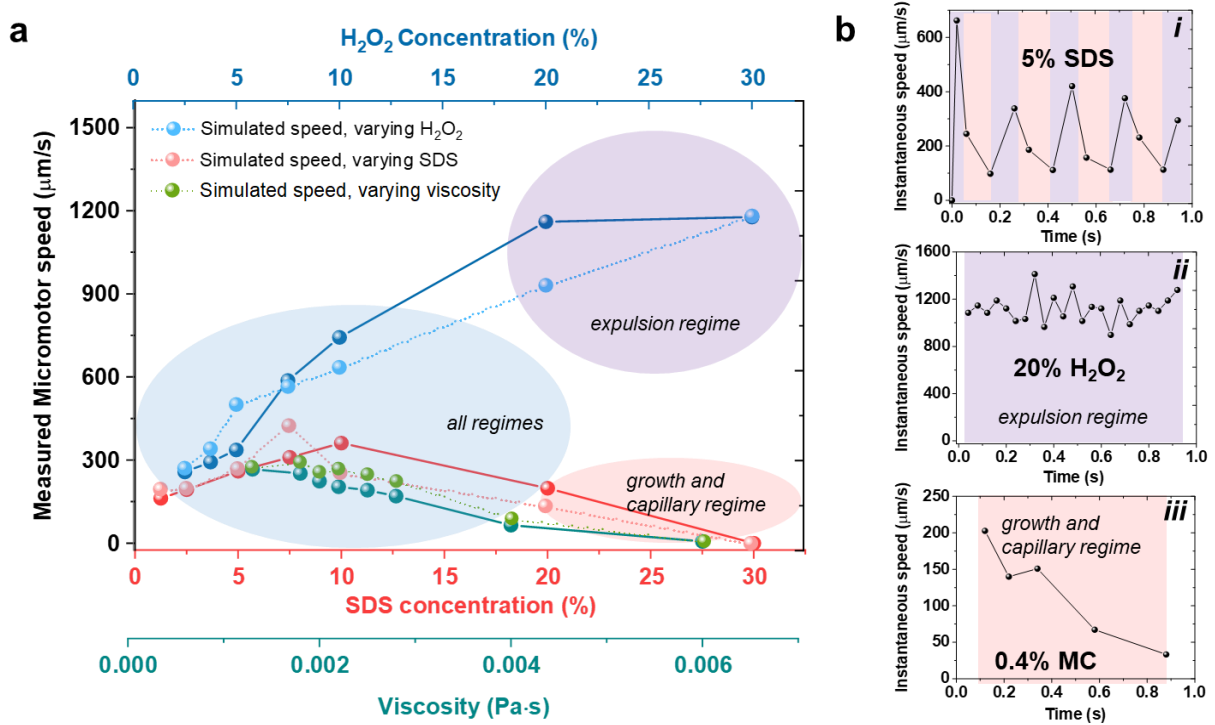

**Figure 5**. a) Graph showing measured and simulated average speeds of micromotors depending on the $H_2O_2$, SDS concentrations and the medium viscosity. During the experiments, switching of propulsion mechanisms is observed, which is controlled by the fuel composition. The expulsion regime is dominant for $H_2O_2$ concentrations above 20%, which is illustrated by the violet area. The SDS concentrations of 20% and higher as well as the viscosities of 4mPa·s and higher lead to the dominance of the growth and capillary regimes (indicated by the pink area). All three regimes coexist for the remaining values of the fuel parameters, shown by the blue area. b) Exemplary plots showing the instantaneous speed of micromotors in the 5% SDS solution [at 2.5% $H_2O_2$ and 0% MC] (i), 20% $H_2O_2$ solution [at 5% SDS and 0% MC] (ii), and the 0.4% MC solution [at 2.5% $H_2O_2$ and 5% SDS] (iii).

## 3. Conclusion

The behavior of 50 µm long catalytic conical micromotors in the presence of different concentrations of $H_2O_2$, SDS and MC was investigated. A physical explanation of the bubble-propelled micromotor motion was performed using the established capillary force, bubble growth, and jet-like propulsion mechanisms. Switching of propulsion mechanisms is for the first time unveiled at certain values of the $H_2O_2$, MC and SDS concentrations. In concrete terms, for $H_2O_2$ above 20%, the jet-like propulsion mechanism was dominant, while for viscosities above 4 mPa·s, as well as, SDS concentrations above 20%, the bubble growth mechanism was the most contributing mechanisms for the micromotor speed. The capillary regime always contributed as the bubble had to move through the tube to grow and to be expelled. Besides, it was observed that for certain experimental parameters all three propulsion mechanisms performed a concerted action. In particular, for $H_2O_2$ concentration at 2.5%, 3.75%, 5%, 7.5% and 10%, for the MC at 0.05%, 0.1%, 0.15%, 0.2%, 0.25%, and for SDS at 1.25%, 2.5%, 5%, 7.5% and 10%, a coexistence of all three propulsion mechanisms was substantiated according to our experiments. Thus, a theoretical model was proposed to calculate the speed for the bubble-propelled micromotors, including the contributions of the three known-propulsion mechanisms, resulting in good agreement with the obtained experimental data. This offers new possibilities to optimize and predict the performance of bubble-propelled micromotors for different applications in various working environments, including new applications such as in-situ sensing of the medium viscosity, as well as for fundamental understanding of the impact of various geometrical parameters and experimental conditions (length, semi-cone angle, surface properties) on the propulsion of conical catalytic micromotors. In future, other scalability and geometrical parameters and surface properties will be considered to fine-tune the model.

**4. Experimental Section**

*3D structures design and writing:* The micromotors used were made by two-photon laser lithography (Photonic Professional from Nanoscribe). The program to control the laser polymerizing the photoresist (IP-Dip of Nanoscribe) was created in the software Describe of Nanoscribe. Briefly, the photoresist was poured onto a quartz substrate and then the writing of the microtubes was done using galvanostatic mode [16] After structuring the micromotors using 3D laser lithography, the solvent was subsequently removed by using a critical point dryer (CPD), to avoid the collapse of the microstructures due to the abrupt surface tension. Then, the structures were coated with Cr (10 nm), Ni (130 nm), Ti (20 nm) and Pt (30 nm) by e-beam evaporation (Edwards E-Beam). It is important that platinum must form the outer layer, otherwise no chemical reaction is possible. Although Pt layer is also present in the outer part of the microtube, one would expect some catalytic reactions which lead to diffusiophoresis motion outside but, considering the size of the micromotors, this effect on the speed and performance can be neglected.

*Fluid preparation:* in order to prepare the liquids, 15 ml Eppendorf tubes were filled with 5 ml deionized (DI) water, for experiments with varying SDS and MC concentrations, 500 µl of 30%

$H_2O_2$ was added to get a final concentration of 2.5% $H_2O_2$. Subsequently, SDS (sodium dodecyl sulphate from Sigma-Aldrich Pty Ltd) was added to the solution (5 mg SDS per 100 ml fluid). This concentration was constant for experiments with varying $H_2O_2$ and MC concentrations. It was changed from 1.25 wt% to 30 wt% for experiments with varying surfactant concentration, when the $H_2O_2$ concentration was kept constant at 2.5 wt%. Then, MC (methylcellulose from Sigma-Aldrich Pty Ltd) was added in the concentrations from 0.05 wt% (0.05 mg per 100 ml fluid) to 0.6% wt for observing the influence of a more viscous media. It is important to ensure that the liquid was first added to the Eppendorf tube and MC was added only at the end, otherwise clumping would occur as the MC is powder. The solution consisting of DI-water, $H_2O_2$, SDS and MC was then immersed in the ultrasonic bath for 30 minutes to achieve a complete solution. Since $H_2O_2$ is highly reactive, the Eppendorf tubes should be carefully closed, and the lid additionally sealed with Parafilm to ensure a longer storage time.

*Microfluidic platform fabrication:* A Parafilm chip was used to carry out the experiments. For this purpose, a conventional glass slide and a cover glass are first placed in acetone and cleaned for 3 minutes with an ultrasonic bath. This process is repeated again with isopropanol. The residues are removed with a nitrogen gun. Parafilm is cut into strips about 3 mm wide. The strips should be about as long as the shortest side of the cover glass. After removal of the protective layer, the strips of Parafilm were stuck onto the cover glass with the help of a tweezer. Further strips were attached to the cover glass with some distance until 3 chambers were formed, making sure that there were no bubbles under the strip, otherwise the liquid could run under it and loosen the strip. Afterwards two more layers of Parafilm were applied to make the chip higher and to ensure an unhindered movement of the micromotors. The height of the chip can be varied by the number of strips of Parafilm on top of each other. The Parafilm+cover glass structure was then placed on the cleaned glass slide. The resulting microfluidic platform with three chambers was heated on the heating plate to 120 °C for about 5 minutes until the Parafilm melted, which was visible by a glassy coloration. By applying a slight pressure with tweezers, a complete connection between Parafilm and the glass slide and cover glass was secured. The finished chips were allowed to cool down to room temperature before use.

*Video recording and analysis:* videos were recorded with a high-speed camera (Phantom) at 50 frames per second. The videos were analysed using ImageJ (open source post-processing software). The MTrackJ plugin was used to determine the speeds, which enables manual tracking of the micromotors. In addition to the speed, the bubble diameter was also determined manually using the measurement function built into ImageJ. The bubble release frequency was determined manually by reviewing each frame and counting the generated bubbles. The collected data were stored in Excel.

**Supporting Information**

This work was supported by the German Research Foundation SPP 1726 "Microswimmers – From Single Particle Motion to Collection Behavior". O.G.S. acknowledges financial support by the Leibniz Program of the German Research Foundation (SCHM 1298/26-1). V.M.F. acknowledges partial support through the MEPhI Academic Excellence Project (Contract # 02. a03.21.0005). This work is part of the projects that have received funding from the European Research Council (ERC) under the European Union's Horizon 2020 research and innovation program (grant agreement nos. 835268 and 853609).

Received: ((will be filled in by the editorial staff))
Revised: ((will be filled in by the editorial staff))
Published online: ((will be filled in by the editorial staff))